\documentclass[12pt]{article}
\usepackage{scicite}
\usepackage{graphicx}
\usepackage{psfig}
\newenvironment{sciabstract}{%
\begin{quote} \bf}
{\end{quote}}

\newcounter{lastnote}

\newcommand{\dpsr}{PSR J1719$-$1438}

\title{
Transformation of a Star into a Planet
in a Millisecond Pulsar Binary.\\ }
\author
       {M. Bailes$^{1,2,3}$\footnote{To whom correspondence should be addressed. Email: mbailes@swin.edu.au}, 
       S. D. Bates$^{4}$ , V. Bhalerao$^5$, N. D. R. Bhat$^{1,3}$,\\ M. Burgay$^{6}$, S. Burke-Spolaor$^{7}$, 
       N. D'Amico$^{6,9}$, S. Johnston$^{7}$,\\ M. J. Keith$^{7}$, M. Kramer$^{8,4}$, S. R. Kulkarni$^5$, L. Levin$^{1,7}$, 
       A. G. Lyne$^{4}$,\\ S. Milia$^{9,6}$, A. Possenti$^{6}$, L. Spitler$^{1}$,
       B. Stappers$^{4}$, W. van Straten$^{1,3}$\\
\\
$^{1}$Centre for Astrophysics and Supercomputing,\\ Swinburne University of Technology, PO Box 218 Hawthorn, VIC 3122, Australia.\\
$^{2}$Department of Astronomy, University of California, Berkeley, CA, 94720, USA.\\
$^{3}$ARC Centre for All-Sky Astrophysics (CAASTRO).\\
$^{4}$Jodrell Bank Centre for Astrophysics, School of Physics and Astronomy,\\ The University of Manchester, Manchester M13 9PL, UK.\\
$^{5}$Caltech Optical Observatories, California Institute of Technology,\\ MS 249-17, Pasadena, CA 91125, USA.\\
$^{6}$INAF - Osservatorio Astronomico di Cagliari,\\ Poggio dei Pini, 09012 Capoterra, Italy.\\
$^{7}$Australia Telescope National Facility, CSIRO Astronomy and Space Science,\\ P.O. Box 76, Epping NSW 1710, Australia.\\
$^{8}$MPI fuer Radioastronomie, Auf dem Huegel 69,\\ 53121 Bonn, Germany.\\
$^{9}$Dipartimento di Fisica, Universit\`{a} degli Studi di Cagliari,\\ Cittadella Universitaria, 09042 Monserrato (CA), Italy.\\
$^*$ To whom correspondence should be addressed; Email: mbailes@swin.edu.au\\
}

\date{}
\begin{document}
\baselineskip24pt

\maketitle

\begin{sciabstract}
Millisecond pulsars are thought to be neutron stars that have
been spun-up by accretion of matter from a binary companion.
Although most are in binary systems, some 30\%
are solitary, and their origin is therefore mysterious. PSR J1719$-$1438,
a 5.7 ms pulsar, was detected in
a recent survey with the Parkes 64\,m radio telescope. We show that it is in
a binary system with an orbital period of 2.2 h. 
Its companion's mass is near that of Jupiter, but
its minimum density of 23 g cm$^{-3}$ suggests that it may be an ultra-low mass carbon white dwarf.
This system may thus have once been an Ultra Compact Low-Mass
X-ray Binary, where the companion narrowly avoided complete
destruction.
\end{sciabstract}

Radio pulsars are commonly accepted to be neutron stars that are
produced in the supernova explosions of their progenitor stars. They
are thought to be born with rapid rotation speeds ($\sim$50 Hz) but within a
few 100,000 yr slow to longer periods because of the braking torque
induced by their high magnetic field strengths
($\sim$10$^{12}$\,G). By the time their rotation periods have reached
a few seconds the majority have ceased to radiate at radio wavelengths.
The overwhelming majority ($\sim 99\%$) of slow radio pulsars are solitary objects.
In contrast $\sim 70\%$ of the millisecond pulsars (MSPs) are members
of binary systems 
and possess spin frequencies of up to 716 Hz\cite{hrs+06}.
This is consistent with the standard model for their origin in which an otherwise dead
pulsar is spun-up by the accretion of matter from a companion star
as it expands at the end of its life\cite{bv91}. Through some process yet to
be fully understood, the recycling not only spins up the neutron star
but leads to a large reduction of the star's magnetic field strength
to $B\sim$$10^8$\,G and usually leaves behind a white dwarf companion of typically
0.2-0.5 M$_\odot$.
The lack of a compelling model for this reduction of the magnetic field
strength with continuing mass accretion, and issues between the
birthrates of MSPs and their putative progenitors, the low-mass X-ray binaries (LMXBs) 
led to an early suggestion\cite{gb88} that accretion induced collapse of a white dwarf 
might form MSPs ``directly'' in the cores of globular clusters, and possibly in the Galactic disk.

In the standard model, the reason why some MSPs possess white dwarf companions
and others are solitary is unclear. Originally it was proposed that solitary MSPs might
be formed from a different channel, in which a massive ($M>0.7$ M$_\odot$) white dwarf coalesces 
with a neutron star\cite{vb84}. The binary pulsar-white dwarf system PSR J1141$-$6545\cite{klm+00}
is destined to merge in $<$ 2 Gyr and thus is a potential progenitor for this scenario. At lower white dwarf masses,
the final product is less clear, as the mass transfer can stabilise\cite{bv85}. 
From an observational point of view, the 
``black widow'' MSPs may give some insights. In these systems an
MSP is usually accompanied by a low-mass $\sim0.02$-$0.05$\,M$_\odot$ 
companion in close orbits of a few hours. It was initially believed these
systems might evaporate what was left of the donor star\cite{fst88}, but
other examples\cite{sbm+98} meant that the timescales were too long.

The MSP population was further complicated by the detection of an
extra-solar planetary system in orbit around the fifth MSP found in the Galactic disk, 
PSR B1257+12\cite{wf92}. This
system has two $\sim$3 Earth-mass planets in 67- and 98-day orbits, and a
smaller body of lunar mass in a 25\,d orbit.
The planets were probably formed from a disk of material. The origin of this disk is
however the subject of much speculation, ranging from some 
catastrophic event in the binary that may have
recycled the pulsar\cite{heuv92} to ablation\cite{rst92} and supernova fall-back\cite{hsc09}. 
A large number of potential models for the creation of this
system have been proposed, and are summarised in the review by Podsiadlowski\cite{pod93}.
Although more than another 60 MSPs ($P<20$\,ms) have been detected   
in the Galactic disk since PSR B1257+12, until now none have possessed
planetary-mass companions.

\dpsr \,was discovered in the High Time Resolution Universe survey
for pulsars and fast transients\cite{kjv+10}.
This $P=5.7$ ms pulsar was also detected in archival data from the
Swinburne intermediate latitude pulsar survey\cite{ebsb01}. Its mean 20\,cm flux density 
is just 0.2 mJy but at the time of discovery was closer to 0.7 mJy
due to the effects of interstellar scintillation. We soon commenced regular timing of the pulsar
with the Lovell 76-m telescope 
that soon revealed that the pulsar was a member of a binary with an
orbital period of 2.17\,h and a projected semi-major axis  of just $a_{\rm p} \sin i$=1.82 ms (Fig 1).
Since then we have performed regular timing of the pulsar at the Parkes and
Lovell telescopes that have enabled a phase-coherent timing solution over a one year
period. There is no evidence for any statistically significant orbital eccentricity with a formal
2-$\sigma$\, limit of $e<0.06$.

With these observations, we can explore the allowed range of companion masses from the binary mass function
that relates the companion mass $m_{\rm c}$, orbital
inclination angle $i$ and pulsar mass $m_{\rm p}$ to the observed projected pulsar 
semi-major axis $a_{\rm p}$, orbital period $P_{\rm b}$ and
gravitational constant $G$:
\begin{equation}
f(m_c) = {{4\pi^2}\over{G}}{{(a_{\rm p}\sin i)^3}\over{P_{\rm b}^2}}={{(m_c \sin i)^3}\over{(m_c+m_p)^2}}=7.85(1)\times10^{-10} M_\odot
\end{equation}
Only a few MSPs in the Galactic disk have accurate masses\cite{jhb+05, vbv+08,
dpr+10}, and these range from 1.4--2.0 M$_\odot$. 
Assuming an edge-on orbit ($\sin i=1$) and pulsar mass $m_{\rm p}=1.4$ M$_\odot$ 
$m_{\rm c} > 1.15\times 10^{-3}$ M$_\odot$, ie approximately the mass of Jupiter. 

We can 
accurately determine the component separation ($a=a_{\rm p}+a_{\rm c}$) for the PSR J1719--1438 binary 
given the observed range of MSP masses using
Kepler's third law and because, $m_{\rm c}  \ll $ $m_{\rm p}$, to a high degree of accuracy,
\begin{equation}
a =0.95 R_\odot\Big(\frac{m_{\rm p}}{1.4\,M_\odot}\Big)^{1/3}
\end{equation}
making it one of the most compact radio pulsar binaries.
For large mass ratios, the Roche Lobe
radius of the companion\cite{pac71} is well approximated by 
\begin{equation}
R_{\rm L} = 0.462 a\left({{m_{\rm c}}\over{m_{\rm c}+m_{\rm p}}}\right)^{1/3}
\end{equation}
and dictates the maximum dimension of the companion star.
For $i=90^o$ and a 1.4 M$_\odot$ neutron star, the 
minimum $R_{\rm L} =2.8\times 10^4$ km, just 40\% of that
of Jupiter.
On the other hand, for a pulsar mass 
$m_{\rm p} = 2$ M$_\odot$, and $i=18^o$ (the
chance probability of $i>18^o$ is $\sim$95\%), then $R_{\rm L} = 4.2\times 10^4$ km.
 
A lower limit on the density $\rho$ (the so-called mean density-orbital period
relation\cite{fkr85}) can be derived by combining the above equations.
\begin{equation}
\rho = {{3\pi}\over{0.462^3 GP_{\rm b}^2 }} = 23 {\rm \,g\, cm}^{-3}
\end{equation}
This density is independent of the inclination angle and the pulsar mass and
far in excess of that of Jupiter or the other gaseous giant planets whose
densities are $<2$ g cm$^{-3}$. 

The mass, radius and hence the nature of the companion of \dpsr\, are critically
dependent upon the unknown angle of orbital inclination.
After \dpsr, PSR J2241$-$5236\cite{kjr+01} has the smallest mass function of the other binary pulsars
in the Galactic disk, albeit 1000 times larger (Fig 2.). The $\sin^3 i$-dependence of the mass function
could mean that \dpsr\, is a physically similar system, but just viewed face-on. This would
require an inclination angle of just $i=5.7^o$, for which the chance probability is 0.5\%.
The only binary pulsar with a similar orbital and spin period is PSR J2051$-$0827, but the
inclination angle required for mass function equivalence in this case has a chance probability of only 0.1\%.
Of course, as the known population of black widow systems increases, we will eventually 
observe examples of face-on binaries that mimic those with planetary-mass companions.
The current distribution of mass functions among the known population is such that
this is still unlikely.

If the pulsar were energetic and the orbit edge-on, we might hope to
detect orbital modulation of the companion's light curve in the optical because the pulsar heats
the near-side. Our pulsar timing indicates the pulsar's observed 
frequency derivative $\dot \nu\,$ is $-$2.2(2)$\times10^{-16}$ s$^{-2}$, 
not atypical of MSPs.
However $\dot \nu$ is only an upper
limit on the intrinsic frequency derivative\cite{ctk94} ($\dot \nu_{\rm i}$), which is
related to the pulsar's distance $d$ and transverse velocity $V_{\rm T}$ by the
Shklovskii relation.
\begin{equation}
\dot \nu = \dot \nu_{\rm i} - \nu V_{\rm T}^2/(dc)
\label{shk}
\end{equation}
MSPs have relatively high velocities\cite{tsb+99} of 50-200 km s$^{-1}$. At the nominal distance of the pulsar
from its dispersion measure\cite{cl02} (1.2 kpc) it would take an MSP transverse 
velocity of only 100 km s$^{-1}$ for almost all of the observed $\dot \nu$ to be caused
by the proper motion of the pulsar. 

In the case of negligible proper motion, we can derive the most optimistic 
impact of the pulsar's radiation for optical detectability by assuming isotropic
pulsar emission, a companion albedo of unity, that the companion is a blackbody, 
and that the orbit is edge-on, thus
maximising the illuminated region of the companion.
The spin-down energy of a pulsar is $\dot E=-4 \pi^2 I \nu \dot \nu$, where $I$ is the moment
of inertia of a neutron star and $\nu$ is the spin frequency. We
find a maximum effective temperature of 4500\,K and a peak $R$-band magnitude
of 26$-$28, depending upon the assumed 1.2(3) kpc distance to the pulsar\cite{cl02} and
the unknown radius of the companion, which we assume is close to the Roche Lobe radius.

We observed the field surrounding \dpsr\, 
with the Keck 10-m telescope in the $g$, $R$ and $I$ bands
using the LRIS instrument.
If the binary was a face-on
analogue of PSR J2051$-$0827 we might expect to see a star at the location
of the pulsar because the $R$-band magnitude of the binary pulsar companion
in the PSR J2051$-$0827 system is $R\sim$22.5\cite{sklk99} and it is at a
similar distance $d$ from the Sun.
The spin-down luminosity of \dpsr\, is however only 0.4 L$_\odot$ which
is about 30\% of that of PSR J2051$-$0827, and a face-on orbit would
mean only half of the bright side of the companion was ever visible.
This would mean the expected $R$-band magnitude would be reduced to $R\sim24.5$,
however at the position of the pulsar there is no visible companion down
to a 3-sigma limiting magnitude of 
$R$=25.4 (1250s), $g$=24.1 (1000s) and $I$=22.5 (1000s) at the anticipated maximum light,
where the values in parentheses indicate the integration times (Fig 3).
The magnitude limit
would appear to reduce the probability 
that \dpsr\, is an extremely face-on analogue of PSR J2051--0827, with
the caveat that the assumed spin-down energy of the pulsar is still an upper limit
because of equation \ref{shk}.

We now consider the more statistically likely possibility that the orbit
is nearly edge-on. In this case the relative velocity of the two constituents is 
$>$ 500 km s$^{-1}$ and could potentially lead to a solid-body eclipse for 
60\,s or so, or if the companion was being ablated we might see excess
dispersive delays at orbital phase 0.25 when the pulsar is on the far side of the companion at 
superior conjunction. Ordinarily the 20\,cm mean flux density of 0.2 mJy would
make these effects difficult to detect, but a bright scintillation band occurred
during one of our long integrations on the source, increasing the flux density sufficiently for us to
assert that there are no excess delays or solid-body eclipses occurring in the system (Fig 1). 
The extremely
small dimension of the Roche lobe of the companion only precludes inclination angles of $i>87^o$. 
Inspection of Fig 2 shows that it is completely impossible to fit a hydrogen-rich planet
such as Jupiter into the Roche Lobe of the planetary-mass companion. 
Although difficult, He white dwarfs might just fit if the computational models\cite{db03} are 
slighty in error ($\sim10$\%), or the orbit is moderately face on. A carbon white dwarf on the 
other hand can easily fit inside the Roche Lobe for any assumed inclination angle. We thus conclude that the
companion star(planet) is likely to be the remains of the degenerate
core of the star that recycled the pulsar, and probably comprised of He or heavier elements
such as carbon. 

In the standard model, this MSP would have been spun-up
by the transfer of matter from a nearby companion star to near
its current period. 
The UC LMXBs such as
XTE J0929$-$314\cite{gcmr02} are good potential progenitors of \dpsr. 
These systems have orbital
periods of tens of minutes and higher ($\sim10\times$) companion masses. They
have also been found to exhibit Ne and O lines in their spectra\cite{jpc01}, suggesting
that their companions are not He white dwarfs.
Importantly, their spin periods are comparable to that of \dpsr.
As matter is transferred from the degenerate companions to the neutron star, the orbits widen and
the radius of the white dwarf expands due to the inverse mass-radius
relationship for degenerate objects. Deloye and Bildsten\cite{db03} predicted
how the known UC LMXBs would evolve in the future. They demonstrated
that the UC LMXB companions could be comprised of either He or
Carbon white dwarfs and after 5-10 Gyr might be expected to end up as binary pulsars
with orbital periods of $\sim$1.5h. 
 
If \dpsr\, was once a UC
LMXB, mass transfer would have ceased when the radius of the
white dwarf became less than that of the Roche Lobe due to mass loss and out-spiral. 
In the models by Deloye and Bildsten, the He white dwarfs deviate 
from the $M^{-1/3}$ law very near to the Roche Lobe radius and approximate 
mass of our companion star for an edge-on orbit. 
On the other hand, another mass-radius 
relationship for Carbon white dwarfs of very low mass\cite{las91} suggests that it is Carbon white dwarfs that 
have $dR/dM\sim0$ near $M=0.0025$M$_\odot$ (Fig 2).
It thus seems difficult to unambiguously determine the nature of the pulsar companion,
but a scenario in which \dpsr\, evolved from a Carbon white dwarf in an UC LMXB has many attractive
features. It explains the compact nature of the companion, the spin period of
the pulsar and the longer orbital period due to spiral out as a consequence of the
mass transfer that spun-up the pulsar. \dpsr\, might therefore be the descendent of an UC LMXB.

However,
the question still remains: why are some MSPs solitary while others retain white dwarf
companions, and some, like \dpsr\, have exotic companions of planetary mass that are
possibly carbon rich?
We suggest that the ultimate fate of the binary is determined by the mass and orbital period of the donor
star at the time of mass transfer. Giants with evolved cores that
feed the neutron star at a safe ($d> $few R$_\odot$) distance leave behind white dwarfs
of varying mass in circular orbits, with a tendency for the heavier
white dwarfs to be accompanied by pulsars with longer pulsar spin periods ($P>$10 ms). 
Close systems that transfer matter before a substantial core has formed might be
responsible for the black-widow MSPs.
A subset of the LMXBs are driven by gravitational radiation losses, and form the ultra-compact
systems during a second stage of mass transfer. 
Their fate is determined by their white dwarf mass and chemical composition at
the beginning of this phase.
High mass white dwarfs do not overflow their Roche Lobes
until they are very close to the neutron star with orbital periods
of a few minutes. If the orbit cannot
widen fast enough to stop runaway mass transfer we will be left
with a solitary MSP or possibly an MSP with a disk that
subsequently forms a planetary system. 
Low mass white dwarf donors transfer matter at longer orbital periods and naturally
cease Roche Lobe overflow near the current orbital period and
implied mass of the companion of \dpsr. The rarity of MSPs with
planetary-mass companions means that the production of planets is the
exception rather than the rule, and requires special circumstances,
like some unusual combination of white dwarf mass and chemical
composition. 

PSR J1719$-$1438 demonstrates that special circumstances can 
conspire during binary pulsar evolution that allows neutron star stellar
companions to be transformed into exotic planets unlike those
likely to be found anywhere else in the Universe. The chemical
composition, pressure and dimensions of the companion make
it certain to be crystallized (ie diamond). 

\vfill\eject

Table 1. Pulsar ephemeris and derived parameters.

\begin{tabular}{ll}
\hline
Parameter & Value \\
\hline
Right Ascension (J2000)	(hh:mm:ss)		&  17:19:10.0730(1)\\
Declination (J2000)		(dd:mm:ss)		& $-$14:38:00.96(2) \\
$\nu$ (s$^{-1}$) &172.70704459860(3) Hz\\
$\dot \nu$ (s$^{-2}$) & --2.2(2)$\times$10$^{-16}$\\
Period Epoch (MJD) & 55411.0\\
DM (pc cm$^{-3}$) & 36.766(2)\\
$P_{\rm b}$ (d) & 0.090706293(2)\\
$a_{\rm p} \sin i$ (lt-s) & 0.001819(1)\\
$T_0$ (MJD) & 55235.51652439 \\
$e$ & $< 0.06$ \\ 
Data Span (MJD) & 55236-55586 \\
Weighted RMS residual ($\mu$s) & 15 \\
Points in fit & 343 \\
Mean 0.73 GHz Flux Density (mJy) & 0.8$^*$ \\
Mean 1.4 GHz Flux Density (mJy) & 0.2 \\
\hline
Derived parameters & \\
\hline
Characteristic Age (Gyr) & $>$12.5 \\
B (G) & $<$2$\times 10^8$ \\
Dispersion Measure Distance (kpc) & 1.2 (3) \\
Spin-down Luminosity L$_\odot$ & $<$0.40(4) \\
\hline
\end{tabular}
\\
$^*$ Derived from a single observation.

\clearpage

\bibliographystyle{Science}
\bibliography{journals,psrrefs}

{{\bf Note.} Differencing of two summed images
at the expected maximum and minimum light of the companion
also failed to reveal any modulation of flux from any potential 
candidates near the nominal pulsar position.}

{\bf Acknowledgements.}

The Parkes Observatory is part of the Australia Telescope 
which is funded by the Commonwealth of Australia for
operation as a National Facility managed by CSIRO. 
This project is supported by the ARC
Programmes under grants DP0985270, DP1094370 \& CE110001020. 
Access to the Lovell telescope
is funded through an STFC rolling grant. Keck telescope time is
made available through a special collaborative program between
Swinburne University of Technology and Caltech.
We are grateful to J Roy and Y Gupta for early attempts to
obtain a pulsar position with the GMRT.

\vfill\eject

\begin{figure}
\includegraphics[width=0.8\textwidth,angle=-90]{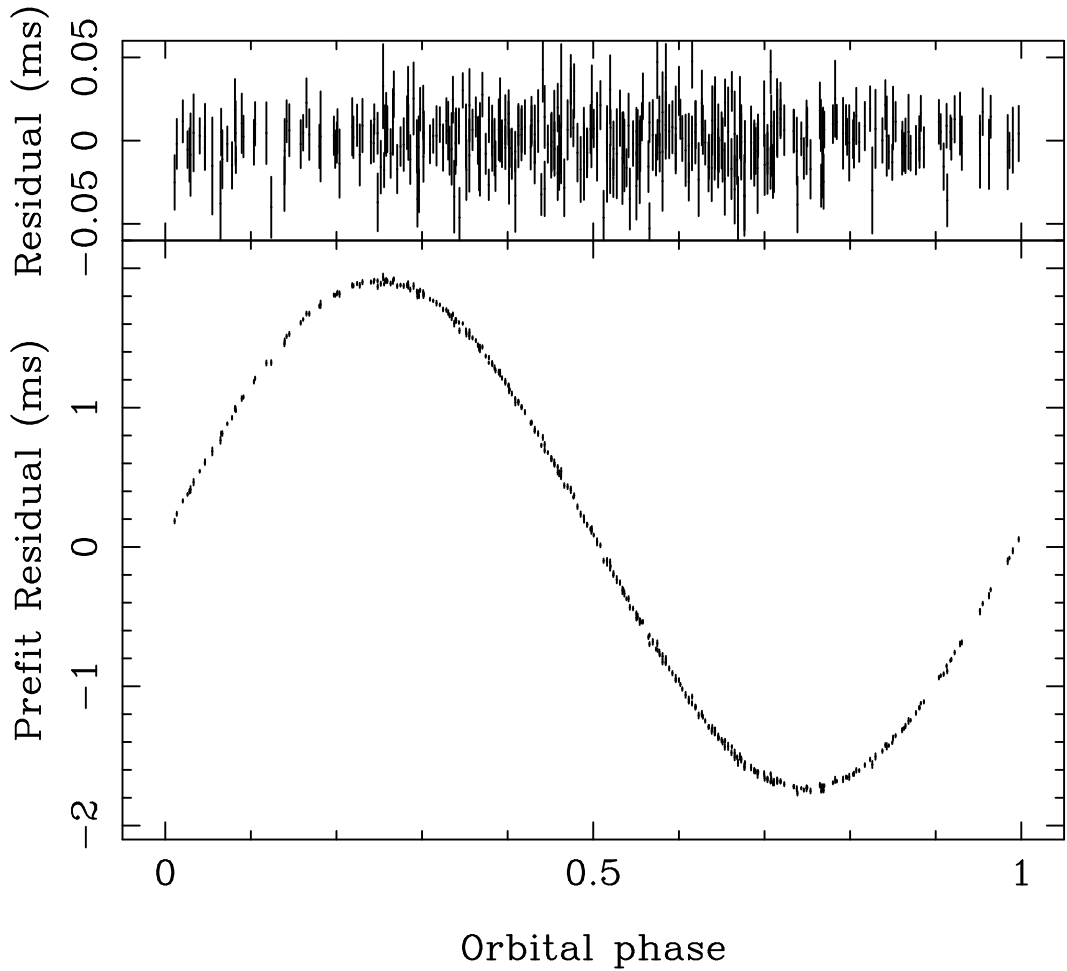}
\caption{
Upper panel: Pulse timing residuals for \dpsr\, as a function of orbital phase 
using the ephemeris in Table 1. Lower panel:  Residuals after setting the semi-major
axis to zero to demonstrate the effect of the binary motion.
There is no significant orbital eccentricity. At superior conjunction 
(orbital phase 0.25) there is no evidence for solid-body eclipses
or excess dispersive delays. The arrival times and ephemeris are
provided in the supporting online material.
}
\end{figure}

\begin{figure}
\includegraphics[width=0.8\textwidth,angle=-90]{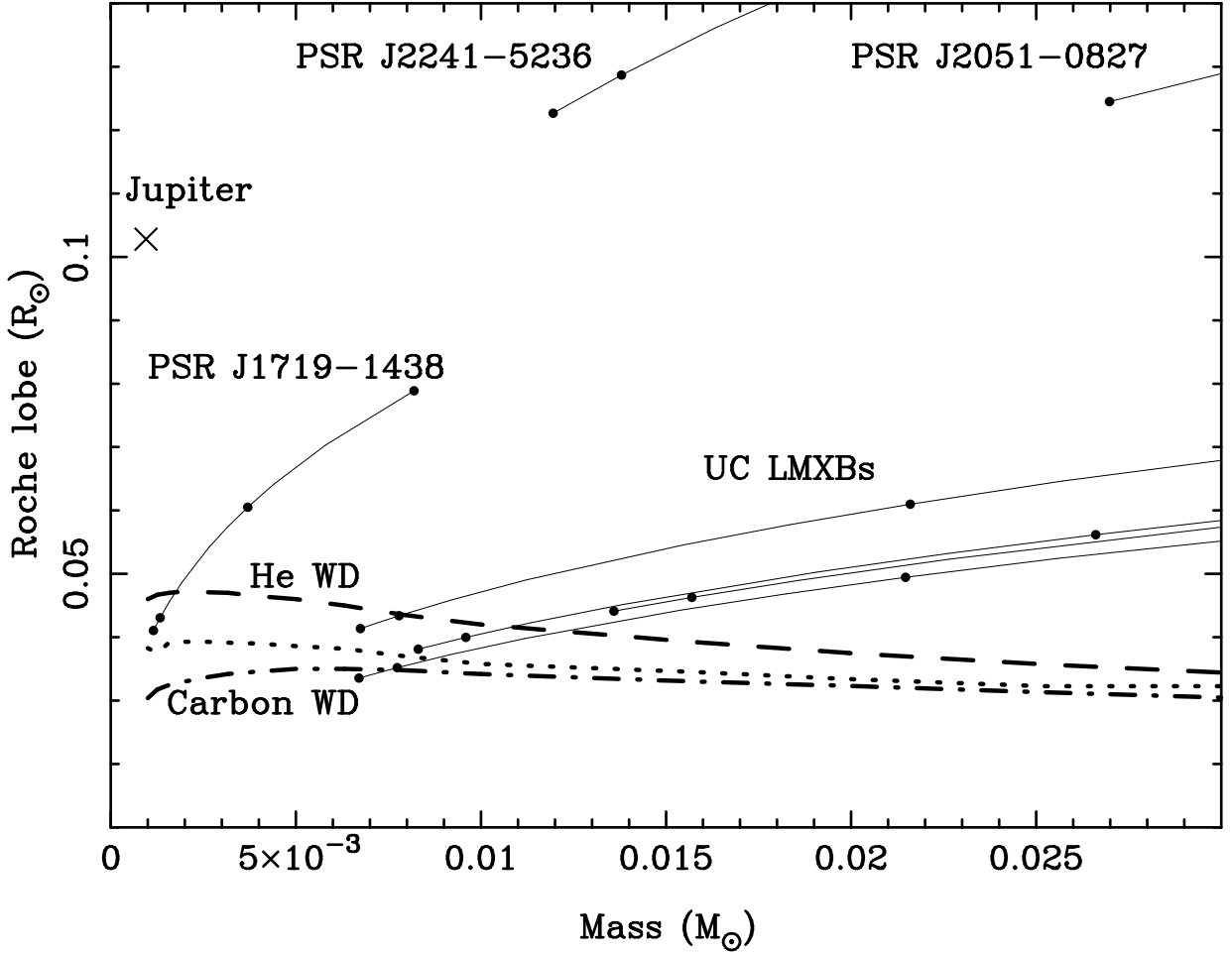}
\caption{The locus of the companion mass and Roche Lobe radii for \dpsr, 
selected ultra-compact LMXBs and black widow millisecond pulsars for different
assumed orbital inclinations. The minimum
companion mass and Roche Lobe radii correspond to $i=90^o$ and a pulsar
mass of 1.4 M$_\odot$. As the unknown angle of inclination decreases, the
companion mass and radius increase, becoming increasingly improbable. 
The bullets from lowest to highest mass represent the minimum ($i=90^o$), median($i=60^o$),
5\% and 1\% a priori probabilities that a randomly-oriented inclination would result in
the mass and radii at least as high as that indicated. The zero-temperature mass-radius relations from Deloye
and Bildsten (2003)\cite{db03} are also shown for low-mass He and Carbon white dwarfs. The dotted line
represents the mass-radius relation for low-mass Carbon white dwarfs computed by Lai, Abrahams
\& Shapiro (1991)\cite{las91}. 
For reference
the mass and radius of Jupiter is shown with an X.
}
\end{figure}

\begin{figure}
\includegraphics[width=0.8\textwidth,angle=0]{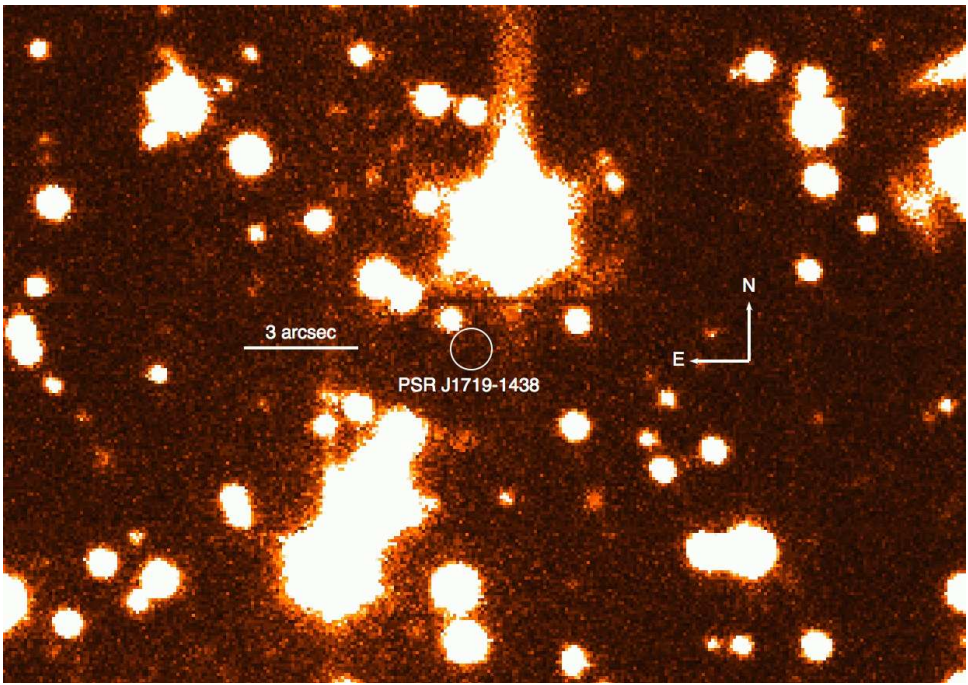}
\caption{
Keck/LRIS 20 minute $R$-band image centred on the location of \dpsr.
The image was constructed from 5 exposures taken during the expected
maximum luminosity of the companion in a total integration time of 1200s. 
}
\end{figure}

\vfill\eject

\end{document}